# Spectroscopic and photometric study of the Mira stars SU Camelopardalis and RY Cephei

**David Boyd**
*BAA Variable Star Section, West Challow Observatory, OX12 9TX, UK*
*Email: davidboyd@orion.me.uk*

**Abstract**

Miras are fascinating stars. A kappa-mechanism in their atmosphere drives pulsations which produce changes in their photometric brightness, apparent spectral type and effective temperature. These pulsations also drive the formation of Balmer emission lines in the spectrum. This behaviour can be observed and investigated with small telescopes. We report on a three-year project combining spectroscopy and photometry to analyse the behaviour of Mira stars SU Cam and RY Cep, and describe how their brightness, colour, spectral type, effective temperature and Balmer emission vary over four pulsation cycles.

## 1. Mira stars

Oxygen-rich Miras are pulsating red giant stars with spectral type late K or M and luminosity class III. Mira variables are evolved stars. They begin their thermonuclear lives burning hydrogen in their core on the main sequence of the Hertzsprung-Russell (HR) Diagram. This hydrogen burning produces helium ash which accumulates in the core. Once the hydrogen in the core runs out, the helium-rich core begins to collapse, which heats the hydrogen-rich shell on top of it until the hydrogen ignites in that shell. While hydrogen is burning in this shell, energy is dumped into the outer envelope of the star causing the outer layers to expand and cool, resulting in the star rising to the upper-right on the HR Diagram, the so-called Red Giant Branch. A helium flash at the top of the Red Giant Branch ignites helium burning to start forming carbon and oxygen in the core. Meanwhile hydrogen burning continues in a shell around the core. Eventually helium in the core becomes exhausted. As the star climbs the Asymptotic Giant Branch it contains both helium and hydrogen burning shells surrounding a degenerate core of carbon and oxygen. The star experiences multiple helium and hydrogen shell flashes or thermal pulses separated by many thousands of years.

Meanwhile, a kappa-mechanism of either dust in the atmospheres of these stars (Fleischer et al. 1995; Höfner et al. 1995) or a hydrogen-ionization zone (Querci 1986) just beneath the visible surface is the likely cause of the approximately annual pulsations which we observe, although this is still the subject of debate (Smith et al. 2002). By this stage the star may have expanded to over a hundred times its original size with a very tenuous outer atmosphere. This extended atmosphere makes it difficult to define the radius, which is also changing with time, and the pulsations are continuously driving the loss of gas and dust into the interstellar medium. An informal review of the problems of understanding Mira variables has been given by Wing (1980).

Mira stars are the subset of pulsating giant stars which have visual amplitudes greater than 2.5 magnitudes and pulsation periods of 100-1000 days. They are named after the prototype, omicron Ceti (Mira). The atmospheres of Miras are sufficiently cool that molecules can form, such as TiO in the oxygen-rich Miras. These molecules absorb light in the visual part of the spectrum which would otherwise have escaped from the star. The resulting TiO molecular absorption bands are a prominent feature in the visual spectrum of Miras. During its pulsation cycle the star appears to brighten as more of its light is emitted in the visual and its effective temperature rises causing some of the molecules to dissociate. Then, as the star fades, it becomes cooler, redder and the molecules reform. The change in the strength of the molecular absorption bands as the star pulsates results in a change in its apparent spectral type. A comprehensive review of our knowledge about Mira stars is given in Willson & Marengo (2012).

## 2. Balmer emission lines

The transient appearance of hydrogen Balmer emission lines in the spectra of Mira stars has been known for over a hundred years since early objective prism observations at the Harvard College Observatory (Maury & Pickering 1897). They were observed to appear and grow as the star approached maximum light and decline and disappear as it faded. One of the earliest detailed spectroscopic studies of a Mira star was that of omicron Ceti by Joy (1926). These early spectroscopic observations were made with photographic plates which were relatively insensitive to red light and only able to record efficiently the shorter wavelength Balmer lines. The use of digital devices now enables all Balmer lines in the visual spectrum to be recorded and measured.

The formation of emission lines in the spectra of long-period variables such as Miras was first explained theoretically by Gorbatskii (1961) as being due to a shock wave produced in the atmosphere of the star. During each pulsation cycle the outward pressure of radiation is countered by the inward pressure of gravity creating a shock which propagates radially outwards ionising hydrogen in the atmosphere and driving mass loss. Many authors including Willson (1976), Fox et al. (1984) and Gillet (1988) have discussed the formation of Balmer emission lines in Mira variables in relation to the production of shock waves. In general, previous observational studies have tended to cover only limited parts of the pulsation cycle or of the visual wavelength range.

## 3. The project

In a conversation with Arne Henden at the Society for Astronomical Sciences Symposium in 2017, he suggested that observing the pulsation of Miras stars spectroscopically might yield interesting results. This led to a three-year project combining spectroscopy and photometry, the results of which are presented here. Because I wanted to observe Mira stars both spectroscopically and photometrically, I needed to choose stars which at their brightest were not too bright to measure photometrically and at their faintest were not too faint to observe spectroscopically using the equipment available. These constraints together imposed a V magnitude range of approximately 8 to 15 on the stars chosen. As I also wanted to be able to observe them throughout the year from my observatory at 52°N, this imposed a practical lower declination limit of around +75°. A search of the AAVSO International Variable Star Index (VSX; Watson et al. 2014) revealed 9 stars catalogued as Miras which fulfilled these criteria, although a check of the magnitudes of these stars reported to the AAVSO over the past year indicated that in most cases their current magnitude range was different from that given in VSX.

A further criterion adopted for practical reasons was that the pulsation period should be less than about 300 days so that it would be possible to accumulate data for at least three pulsation periods over a three-year project. I also wanted to be able to observe several consecutive pulsation cycles of both stars on a cadence of 1/20$^{th}$ of their pulsation period, bearing in mind constraints on observing due to weather, without impacting too severely on other ongoing observing projects. In the end I decided to follow two stars which had been under-observed digitally, possibly because they were not on the AAVSO LPV target list, namely SU Cam and RY Cep. It later emerged that they had rather different spectral types which added another interesting dimension to the project.

The project focused on studying how the photometric brightness and colour, the apparent spectral type and effective temperature, and the behaviour of the Balmer emission lines of these two stars varied over each pulsation cycle and from cycle to cycle. Data analysis was performed using Python software developed by the author which made extensive use of the Astropy package (Astropy Collaboration et al. 2018).

## 4. Observations

Spectroscopy was obtained with a 0.28m Schmidt-Cassegrain Telescope (SCT) operating at f/5 equipped with an auto-guided Shelyak LISA slit spectrograph and a SXVR-H694 CCD camera. The slit width was 23µ giving a mean spectral resolving power of ~1000. Spectra were processed with the ISIS spectral analysis software (ISIS; Buil 2021). Spectroscopic images were bias, dark and flat corrected, geometrically corrected, sky background subtracted, spectrum extracted and wavelength calibrated using the integrated ArNe calibration source. They were then corrected for instrumental and atmospheric losses using spectra of a nearby star with a known spectral profile from the MILES library of stellar spectra (Falcón-Barroso et al. 2011) situated as close as possible in airmass to the target star and obtained immediately prior to the target spectra. Typically, 12 five-minute guided integrations were recorded for each spectrum which gave signal-to-noise ratios ranging from ~100 at maximum brightness to ~10 at minimum. Spectra were calibrated in absolute flux in FLAM units using concurrently measured and transformed V magnitudes as described in Boyd (2020). All spectra were submitted to, and are available from, the BAA spectroscopy database (BAA Spectroscopy Database 2021).

Photometry was obtained with a 0.35m SCT operating at f/5 equipped with Astrodon BVRI photometric filters and an SXVR-H9 CCD camera. All photometric observations were made through alternating B and V filters with typically 10 images recorded in each filter. B and V filters were used because B and V magnitudes of comparison stars are available from the AAVSO Photometric All-Sky Survey (APASS; Henden et al. 2021). These photometric observations were made concurrently with recording spectra. All photometric images were bias, dark and flat corrected and instrumental magnitudes obtained by aperture photometry using the software AIP4WIN (Berry & Burnell 2005). An ensemble of 5 nearby comparison stars was used whose B and V magnitudes were obtained from APASS. Instrumental B and V magnitudes were transformed to the Johnson UBV photometric standard using the measured B-V colour index and atmospheric airmass with the algorithm published in (Boyd 2011). Times are recorded as Julian Date (JD). All magnitudes were submitted to, and are available from, the BAA photometry database (BAA Photometry Database 2021).

Figure 1 shows spectra of SU Cam and RY Cep near maximum and minimum with Balmer emission lines marked and including transmission profiles of the B and V filters used. The spectrum of SU Cam at minimum has been amplified to make it more visible. Figure 1 shows that, at maximum, the spectral type of RY Cep is much earlier than that of SU Cam. At our resolution we are not able to see the detailed structure in the emission lines reported in some higher resolution studies. While neither the Hα or Hβ emission lines contribute significantly in the V band, the Hγ and Hδ emission lines do contribute to the flux recorded in the B filter but calculation shows that this is in most cases much less than 10%. The relative strength of the Balmer lines seen in these spectra will be discussed later.

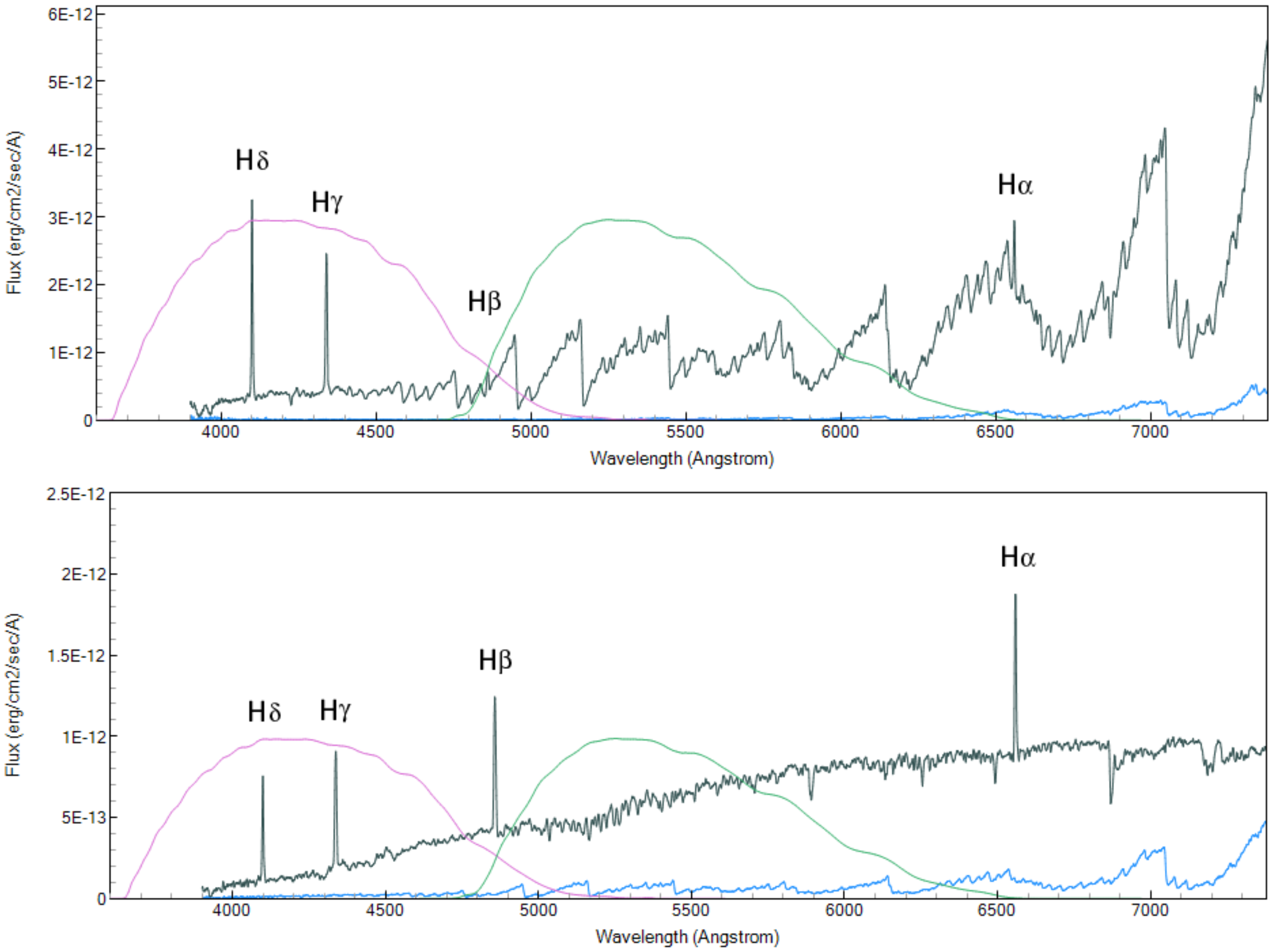

Figure 1. Spectra of SU Cam (upper) & RY Cep (lower) at maximum and minimum brightness with Balmer emission lines marked and including B (red) and V (green) filter transmission profiles. The spectrum of SU Cam at minimum has been amplified 5 times to make it more visible.

The distance reported by Gaia EDR3 (Gaia Collaboration et al. 2016) for SU Cam is 1009 +85 -77 parsec and for RY Cep is 2578 +150 -138 parsec. According to Schlafly & Finkbeiner (2011), the galactic extinction towards SU Cam is E(B-V) = 0.097 and towards RY Cep is E(B-V) = 0.158. Both stars lie out of the plane of the galaxy so the real extinction they experience is likely to be close to these values. We use the formulae in Cardelli et al. (1989) to compute dereddening profiles used to deredden all our spectra.

## 5. Photometric brightness and colour

Photometric B-band and V-band observations of SU Cam and RY Cep cover the period JD 2457994 to 2459341. Figure 2 shows B and V magnitudes and B-V colour index vs JD. Figure 3 shows the V vs B-V colour magnitude diagrams in which each colour-coded cycle traverses a complex loop. The initial point during each pulsation cycle is marked with a larger black dot to make it easier to follow the trajectories in every cycle. The complex path followed during each cycle in Figure 3 is a consequence of the way the flux profile of the spectrum of each star, as integrated by these filters, changes during its pulsation cycle.

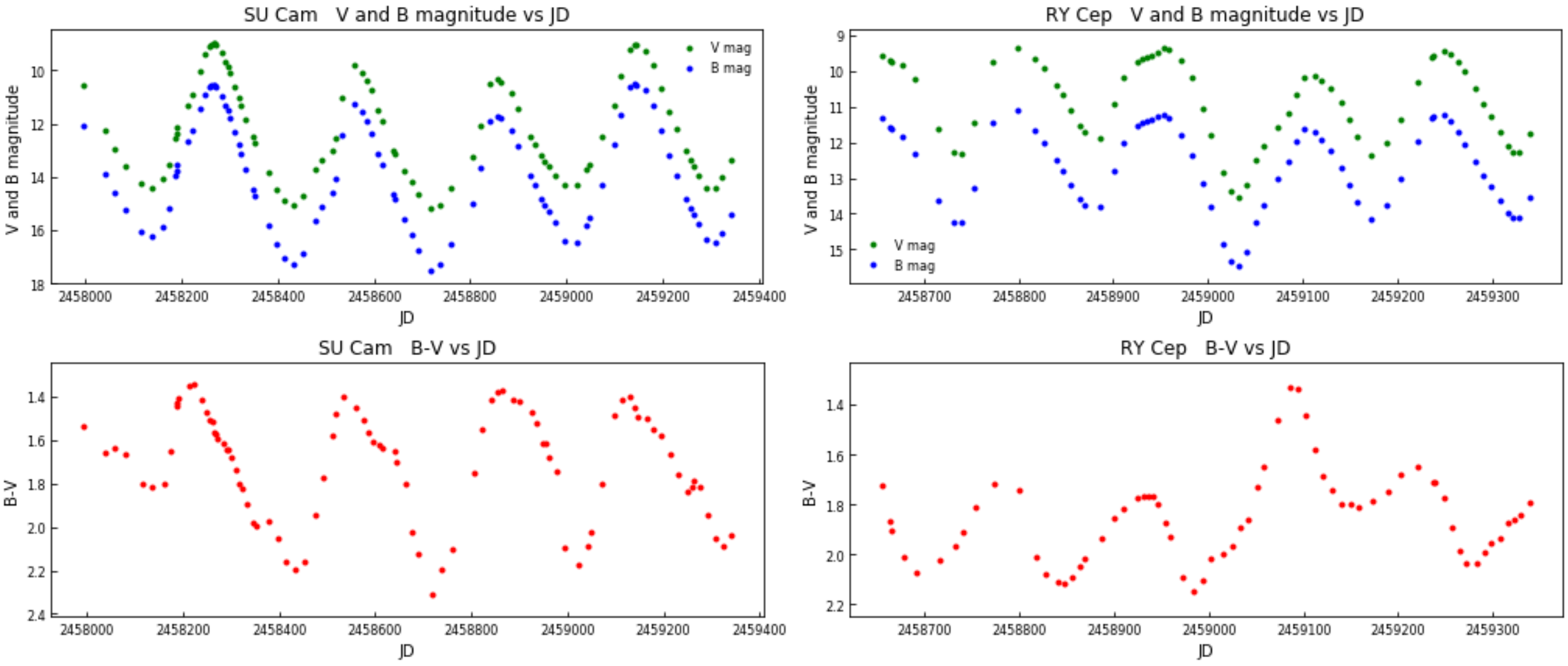

Figure 2. V and B magnitudes and B-V colour index vs JD showing 4 pulsation cycles.

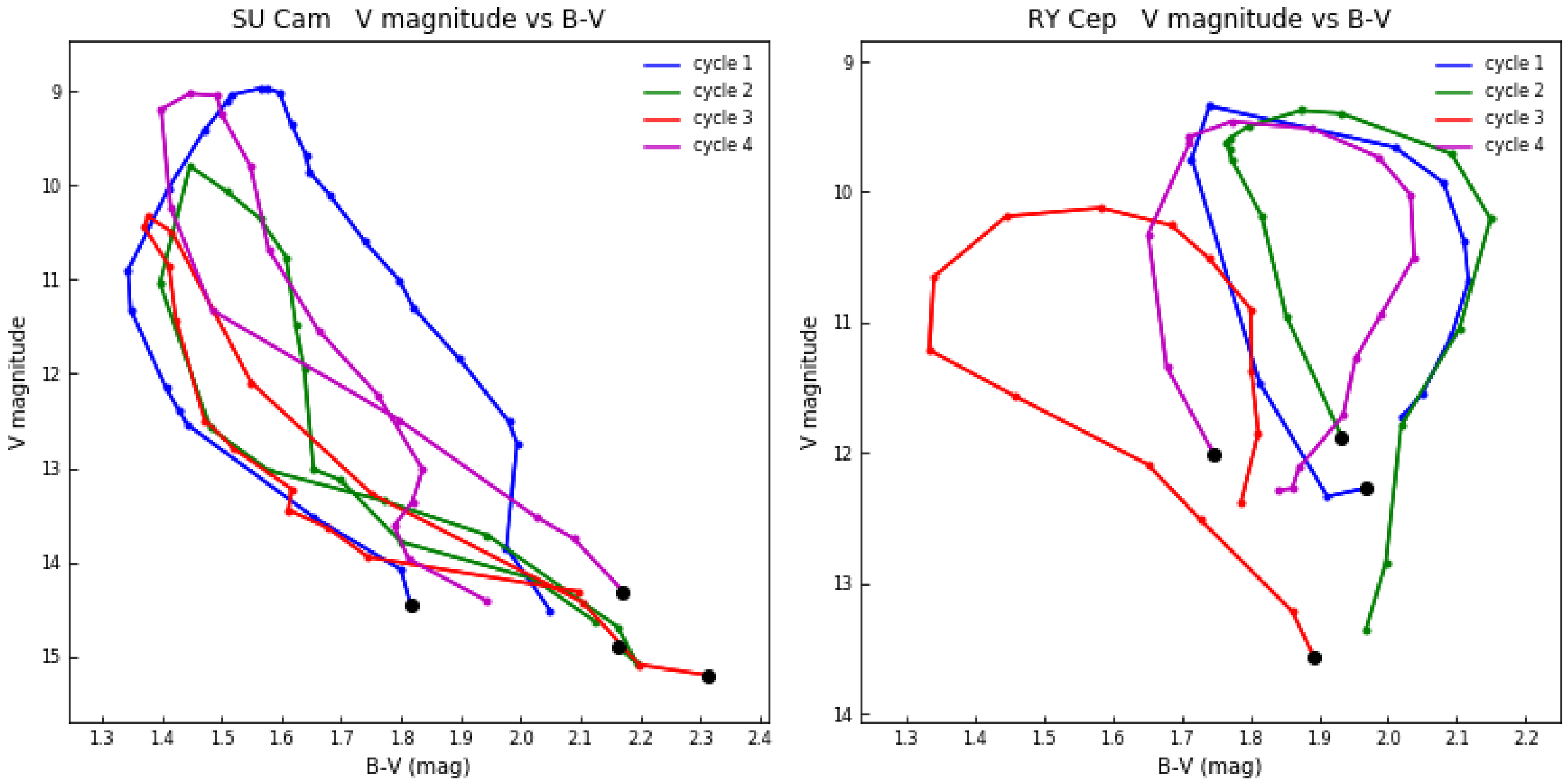

Figure 3. V vs B-V colour magnitude diagrams for each pulsation cycle.

V magnitude measurements around the peak of each of the four recorded pulsation cycles for each star were fitted with a 4$^{th}$ order polynomial in JD and this was used to find the time of maximum magnitude for each cycle. These times were converted to Heliocentric Julian Date (HJD) and used to derive the following linear ephemerides for times of maximum (ToM) with E ranging from 0 to 3:

SU Cam: ToM (HJD) = 2458267.74793(5) + 292.49341(3) * E (1)

RY Cep: ToM (HJD) = 2458798.25577(5) + 152.63684(3) * E (2)

The mean pulsation period of SU Cam over this time interval is 292.49 days and for RY Cep is 152.64 days. The periods currently listed in VSX are 286.25 days and 149.06 days respectively. Using these linear ephemerides, all times were converted to phases of the pulsation cycle with phase 0 occurring at or close to the time of maximum brightness.

B and V magnitudes can be converted to absolute flux using photometric zero points derived from CALSPEC spectrophotometric standard stars (Bohlin, R. C. et al. 2014; STScI 2021). The variation of B and V band flux with pulsation phase for each pulsation cycle is shown for both stars in Figure 4. While there is considerable variation in the profiles from cycle to cycle, it is noticeable that the B and V flux profiles are more sharply peaked in SU Cam in all cycles compared with the broader peaks in RY Cep.

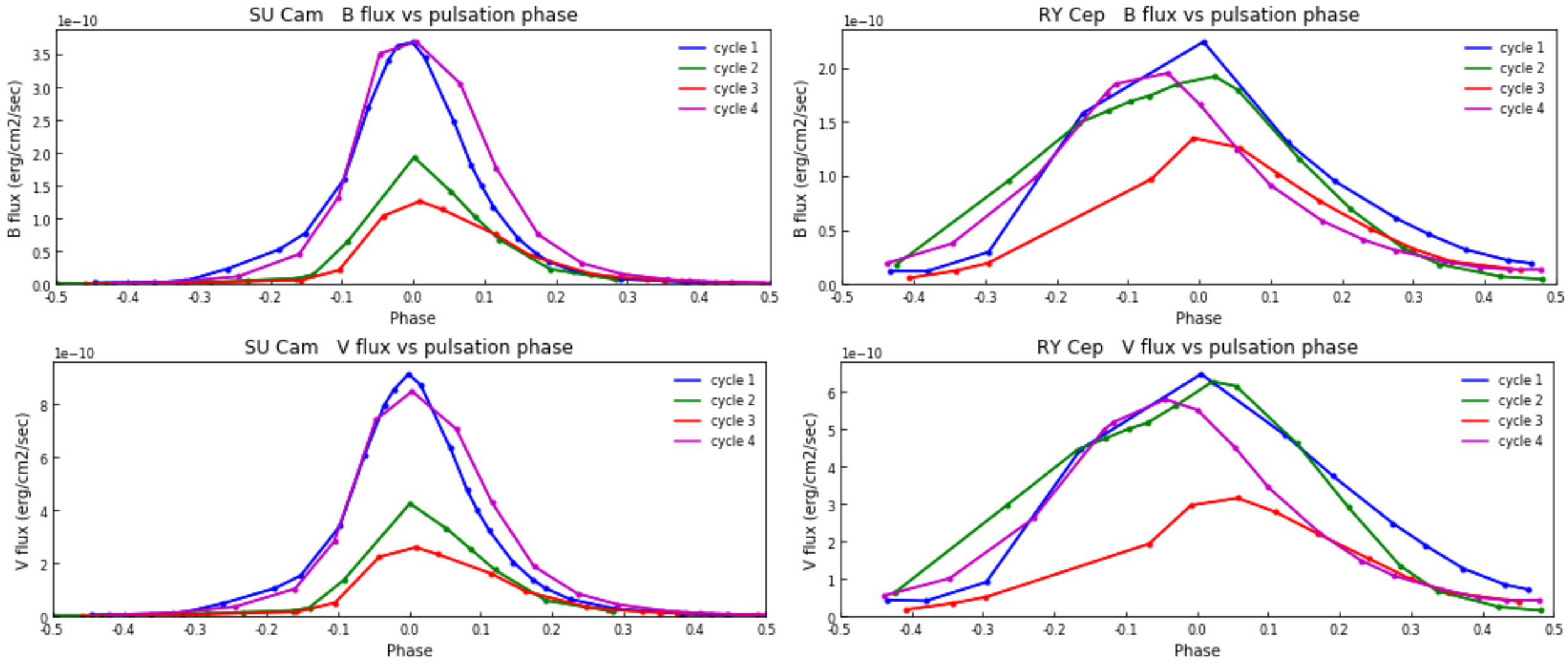

Figure 4. B and V band flux vs pulsation phase for each pulsation cycle.

The reason for the complex behaviour of the B-V colour magnitude diagrams in Figure 3 becomes clearer when we look at the variation of the B-V colour indices over the pulsation phase for each cycle as shown in Figure 5. In most cycles the B-V colour index peaks before maximum brightness then becomes redder as the cycle progresses, thereby traversing a clockwise loop in the colour-magnitude diagram. This effect is more pronounced in RY Cep compared to SU Cam leading to wider loops in the former.

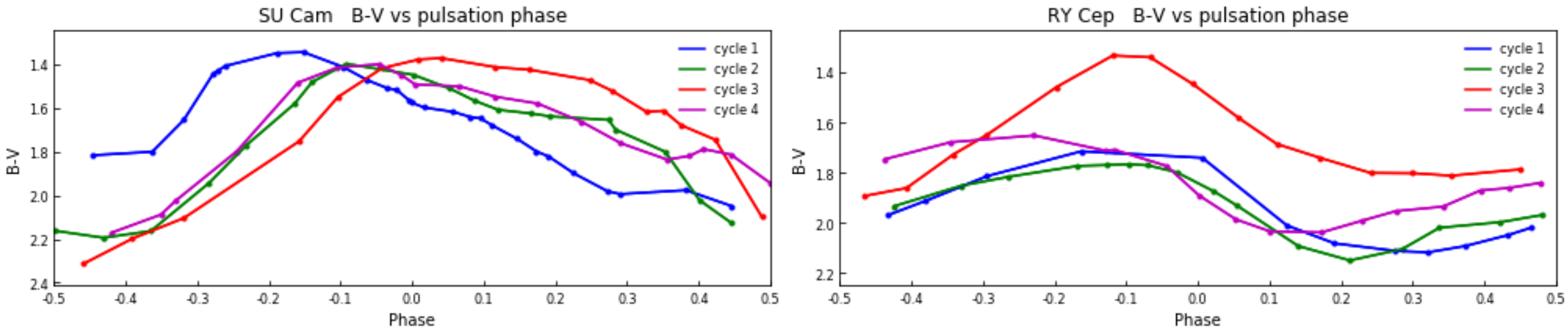

Figure 5. B-V colour index vs pulsation phase for each pulsation cycle.

## 6. Apparent spectral type and effective temperature

Oxygen-rich giant stars of later spectral type exhibit strong TiO molecular absorption bands in their spectra and the strength of these bands is usually taken as an indication of the spectral type of the star. Because the strength of the molecular bands changes over the pulsation cycle in Mira stars, this relationship is more complex. However, for the purpose of our analysis, we will assume that this relationship can be used to assign a spectral type to Mira stars that changes as they pulsate.

Assigning a spectral type to a spectrum is commonly achieved by comparing it morphologically to a range of standard star spectra in the MK spectral classification system and identifying the closest match (Gray & Corbally 2009). MK standard stars available with the MKCLASS stellar spectral classification system (Gray & Corbally 2014) cover the wavelength range 3800 – 5600 Å where atomic absorption lines are concentrated, a legacy of the use of blue-sensitive photographic plates in the early days of the MK standard. In our spectra the flux in this region is relatively low whereas it is considerably stronger towards the red end of the visual range where the molecular bands are prominent. Given our limited spectral resolution and therefore inability to resolve some of the lines in the blue part of the spectrum used for classification, using the full visual range to classify our spectra offers a more practical way of assigning a spectral type.

As all our SU Cam spectra fell within range of the M spectral type, we decided to use the M giant spectra published in Fluks et al. (1994), which are classified on the MK system, to assign an apparent spectral type to each spectrum. The Fluks spectra for spectral types M0 to M10 are defined on the wavelength range 3500 – 10000 Å at an interval of 1 Å. The spectral type of RY Cep became earlier than M as it approached maximum light in each cycle. This meant finding stars with MK standard spectral type K for which we could also obtain spectra. After considering possible sources, we decided to use stars listed in the Perkins catalogue of stars classified on the revised MK system (Keenan & McNeil 1989) for which there are spectra in the MILES library of stellar spectra (Falcón-Barroso et al. 2011). Examination of their spectra showed that they formed a sequence which was both internally consistent and also consistent with the transition to the Fluks M type spectra. The stars used as standard spectral types between K0III and K5III are listed in Table 1.

Prior to using them in our analysis, all spectra being used as standards were interpolated to a wavelength interval of 1 Å and normalised to a mean flux value of unity in the wavelength interval 5610 to 5630 Å which contains no strong spectral features. All our spectra of SU Cam and RY Cep were similarly interpolated to 1 Å and normalised to unity in the same wavelength interval. After removing emission lines each of our spectra was compared with each of the K and M type standard spectra. The differences in flux at each Angstrom between 4000 and 7000 Å were squared and totalled. This gave a quantitative measure of the difference in profile between each of our spectra and each of the standard spectra. For each spectrum there was one spectral type for which this flux difference was a minimum. By fitting a quadratic polynomial to the flux differences around this minimum, it was possible to assign a spectral sub-type to the nearest tenth to each of our spectra. These assigned K and M spectral sub-types are listed for each SU Cam spectrum in Table 2 and for each RY Cep spectrum in Table 3. The spectral types of SU Cam range from M5 to M8 while those of RY Cep range from K4 to M6.

The relationship between effective temperature and spectral type in Miras is not simple and the literature contains a variety of approaches to this problem. After reviewing the options, we decided to adopt a pragmatic approach and use the data on effective temperature and spectral type for K and M giant stars given in van Belle et al. (1999) and apply these to our Mira spectra. We used a polynomial parameterisation of the van Belle data to assign effective temperatures to all our spectra based on their assigned spectral types. These assigned effective temperatures are also listed in Tables 2 and 3.

The variation of effective temperature with pulsation phase over each pulsation cycle is shown in Figure 6. Maximum effective temperature occurs close to the time of maximum brightness in all cycles in both stars. Similar to the B and V flux behaviour in Figure 4, the rise in effective temperature around phase 0 is more narrowly peaked in SU Cam than in RY Cep where the effective temperature changes more gradually through the cycle.

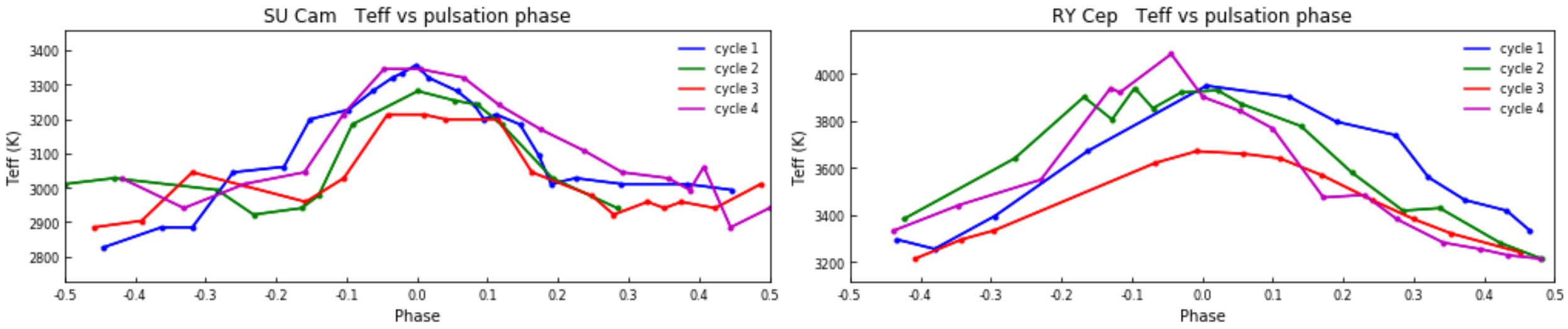

Figure 6. Effective temperature (Teff) vs pulsation phase for each pulsation cycle.

Effective temperatures are plotted against concurrently measured V magnitudes for all SU Cam and RY Cep spectra in Figure 7. Different symbols are used to differentiate the rising and falling branches of each cycle. V magnitude and effective temperature are clearly correlated with both rising and falling branches following the same trajectory. Below 13th magnitude the greater scatter is a consequence of the increasing noise in these spectra. Figure 7 also includes the parameters of quadratic fits to the data including the R-squared value for the correlation. The internal consistency of these plots suggests that our pragmatic approach to assigning effective temperatures was reasonable.

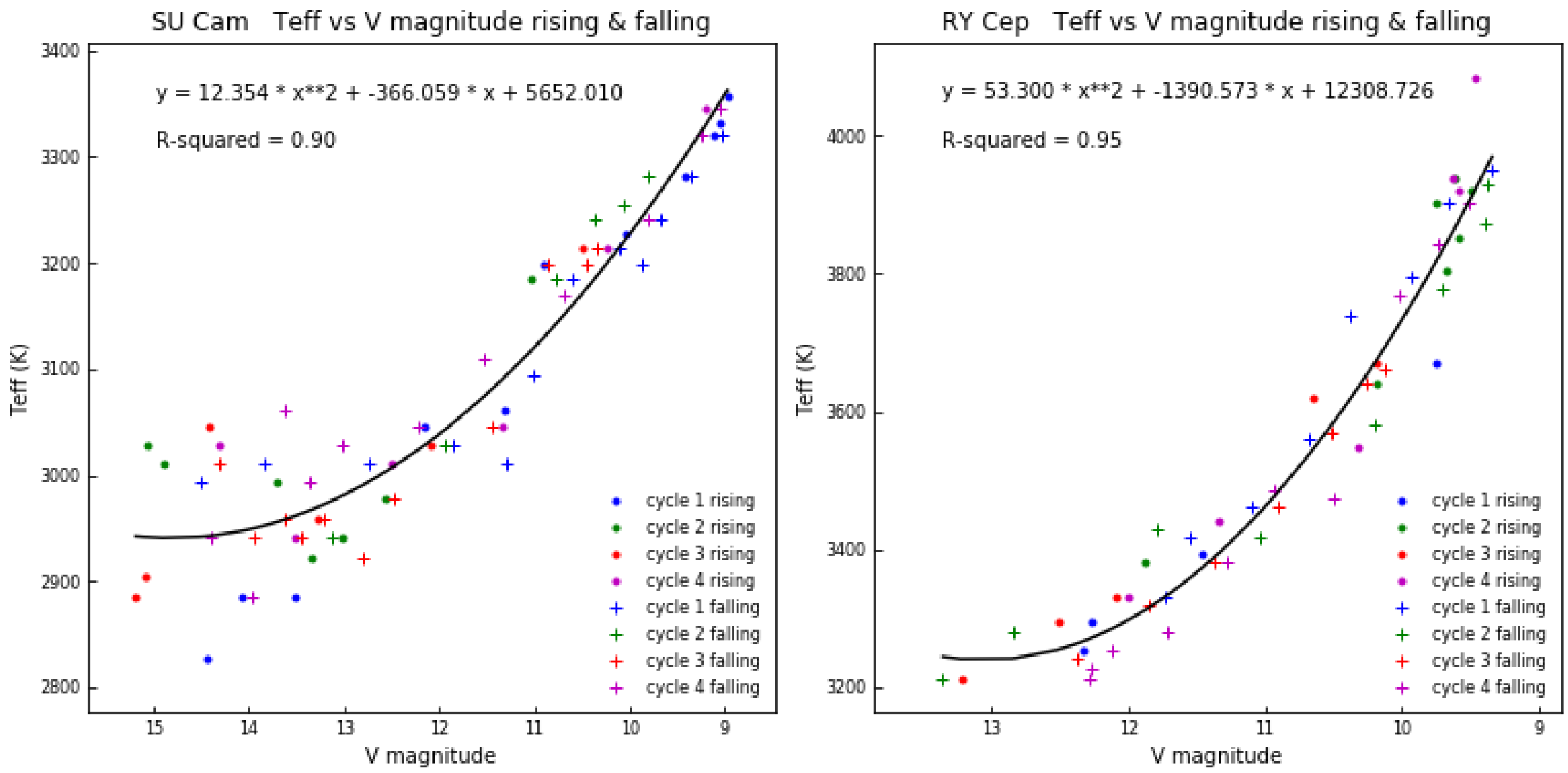

Figure 7. Effective temperature (Teff) vs V magnitude for all spectra plus quadratic fits to the data.

In many stars the B-V colour index serves as a useful proxy for effective temperature. This is not the case in Miras as Figure 8 shows. Different wavelengths probe different depths in their tenuous atmospheres and the varying amounts of molecular material present in the atmosphere during each pulsation cycle change the relative flux in the B and V bands in a complex way which varies during a pulsation cycle and from cycle to cycle. The different path taken by RY Cep during cycle 3 is the result of lower flux in both B and V in this cycle as shown in Figure 4 which results in lower effective temperature. The flux in V in this cycle is also proportionally lower than the flux in B compared to the other cycles giving a bluer B-V colour index as shown in Figure 5. The combined effect is to move the path taken in this cycle down and to the left in the diagram.

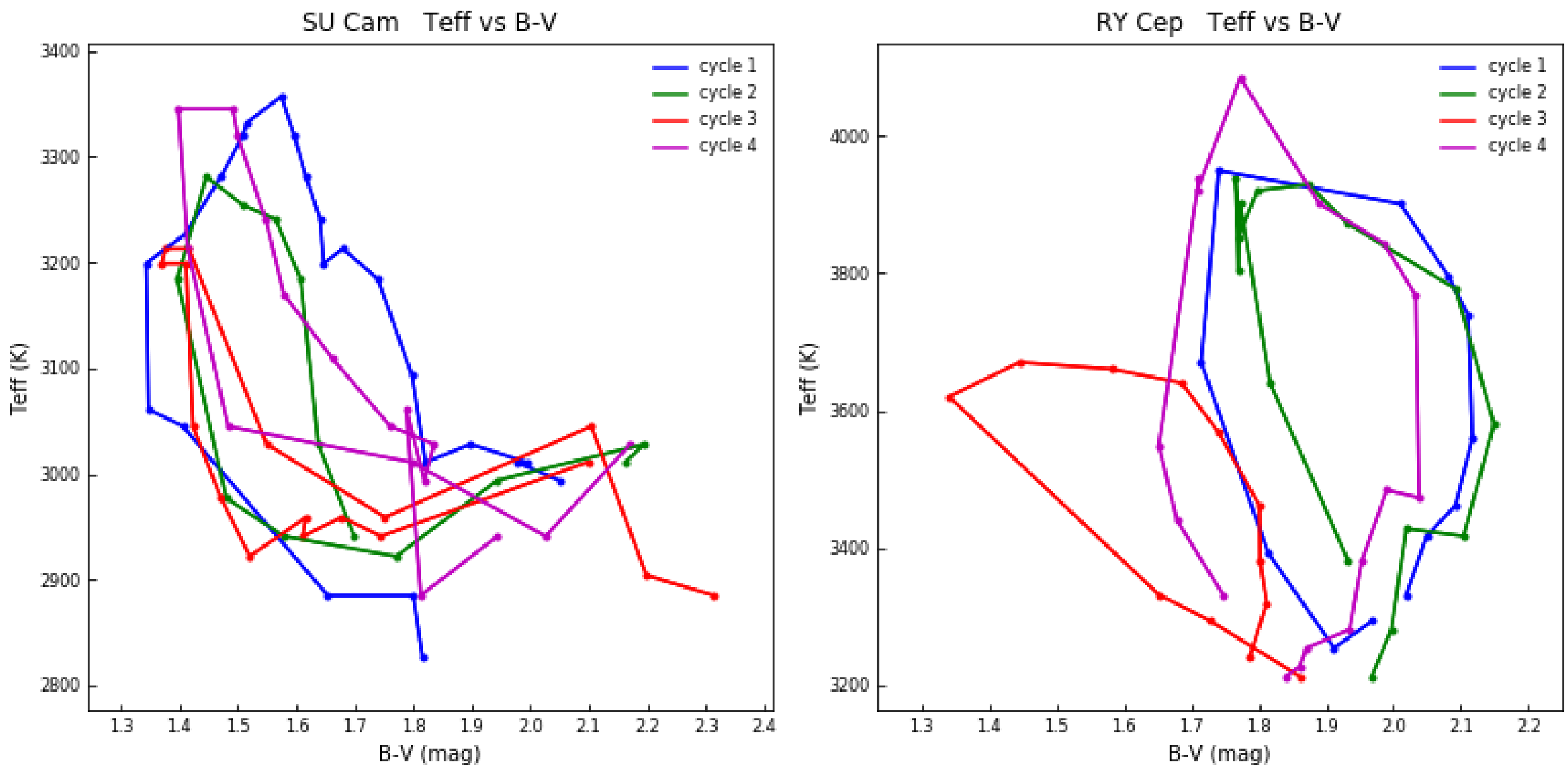

Figure 8. Effective temperature (Teff) vs B-V colour index for each pulsation cycle.

## 7. Balmer emission lines

Shocks produced in the atmosphere of both stars during each pulsation cycle generate emission lines of the hydrogen Balmer series. The strength of these lines tends to increase as the star brightens and decline after it has passed through maximum brightness. As our spectra have been calibrated in absolute flux, the flux in each emission line above the local continuum can be measured using the software PlotSpectra (Lester 2020). These Balmer line fluxes for each spectrum are listed in Tables 2 and 3, while Figure 9 shows how they vary over each pulsation cycle. Where the flux above the local continuum is too small to be reliably measured, the tables contain the value zero. Many emission lines are asymmetric about phase 0 and have considerable skew in their profiles. There is nevertheless a degree of consistency within cycles with all lines peaking around the same phase in the same cycle. There also is a noticeable tendency for emission lines to be broader in phase in RY Cep than in SU Cam.

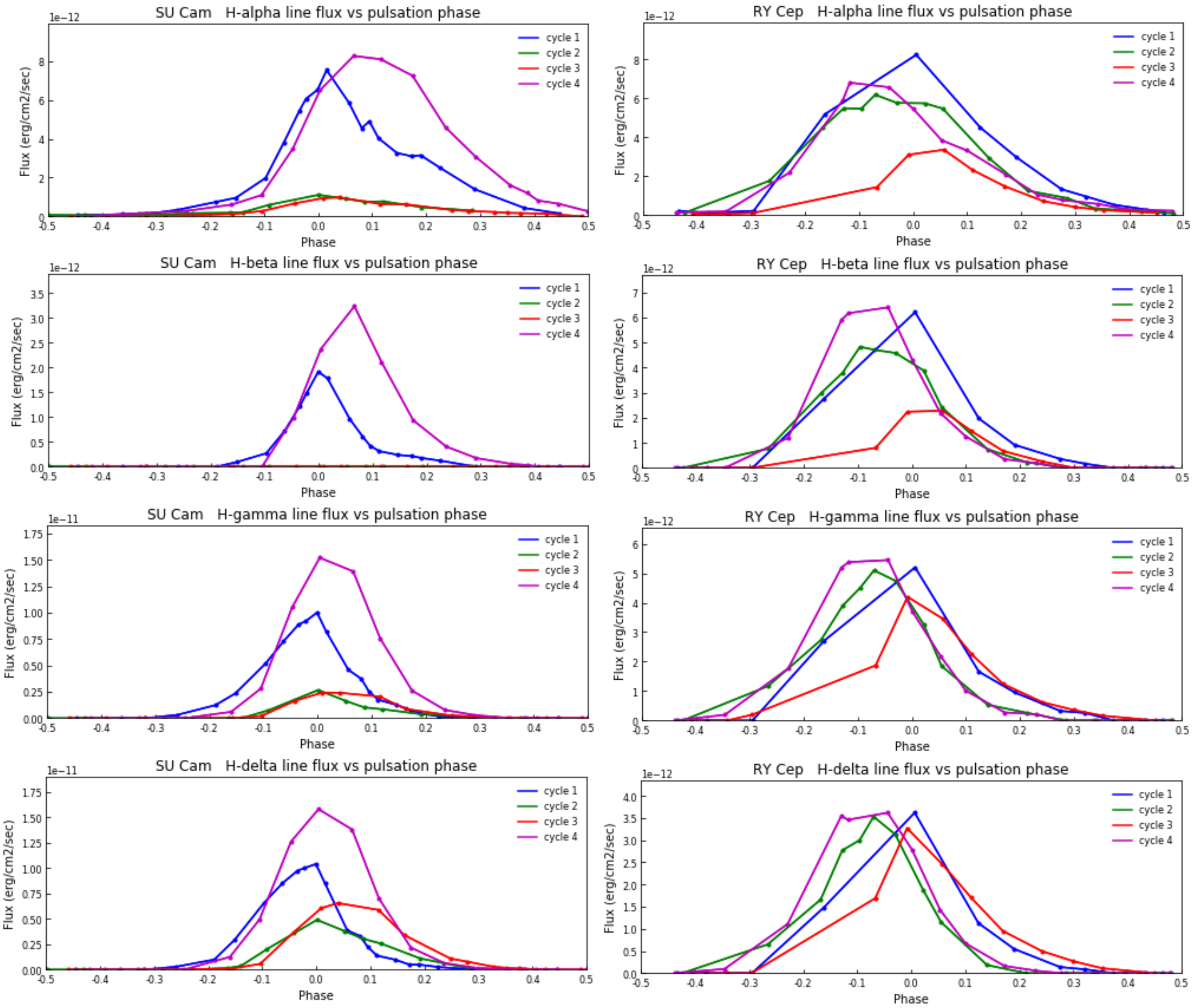

Figure 9. Flux in the Balmer emission lines vs pulsation phase for each pulsation cycle.

Yao et al. (2017) found that, in oxygen-rich Miras, there is a Balmer increment (Hα < Hβ < Hγ < Hδ) for stars with spectral types M5 to M10 whereas there is a Balmer decrement (Hα > Hβ > Hγ > Hδ) for earlier spectral types. They noted that Hβ is sometimes weak which is also our experience. Our spectra shown in Figure 1 are consistent with Yao et al.'s conclusions. To investigate this further we used the Hα/Hδ line flux ratio for both stars as a proxy for the Balmer decrement/increment (>1 = decrement, <1 = increment) and effective temperature as a proxy for spectral type. Figure 10 shows that the Hα/Hδ line flux ratio falls as the effective temperature drops and around 3550 K, equivalent to spectral sub-type M3.6, the Balmer decrement changes to an increment.

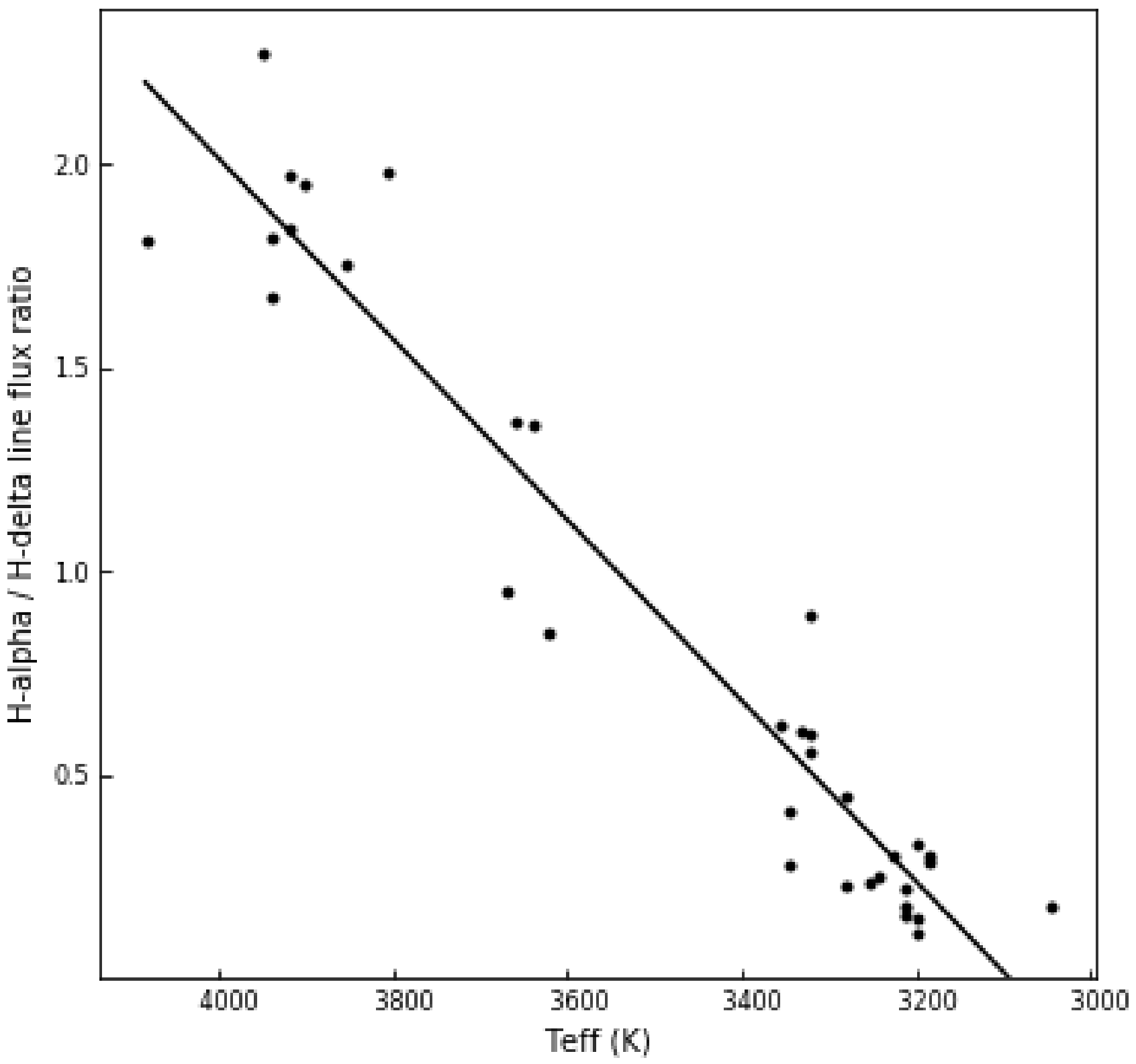

Figure 10. Hα/Hδ line flux ratio vs effective temperature (Teff) plus a linear fit to the data.

Previous studies of Mira stars have analysed the strength and behaviour of Balmer emission lines on a per spectrum basis and have covered only part of their pulsation cycle. Because we have comprehensive coverage of several cycles, we can analyse our data on a per cycle basis. By linearly interpolating and integrating over the emission line profiles we can compute the total flux or energy emitted in each line during each pulsation cycle. Similarly, integrating over the V band profiles in Figure 4 gives a measure of the total energy emitted in the V band in each cycle. These integrated fluxes are listed in Tables 4 and 5. As noted previously (Fox et al. 1984), the shock-induced line flux generally increases with the strength of the pulsation-driven V band flux. Tables 4 and 5 also show that the integrated Balmer line fluxes in each cycle generally follow an increment in SU Cam and a decrement in RY Cep.

## 8. Estimating effective temperature from continuum flux ratios

The spectrum of an oxygen-rich M giant star does not represent its true photospheric continuum because of extensive molecular absorption (Fluks 1994). Wing (1992) developed a technique for characterising the spectra of red variables by making photometric measurements at three wavelengths and deriving an index of TiO band strength which could be used to estimate spectral type.

As we have a set of Mira spectra for which we have computed effective temperatures, we investigated whether a simple variation on this idea could be used to estimate the effective temperatures of Mira stars. We measured the mean flux in the two wavelength ranges 6130 – 6140 Å and 6970 – 6980 Å marked in Figure 11. These are adjacent to TiO molecular band heads and are therefore likely to be regions of the spectrum closest to the true photospheric continuum. Because this involves taking a flux ratio, it does require the spectrum to be calibrated in relative flux across this spectral range but not necessarily in absolute flux.

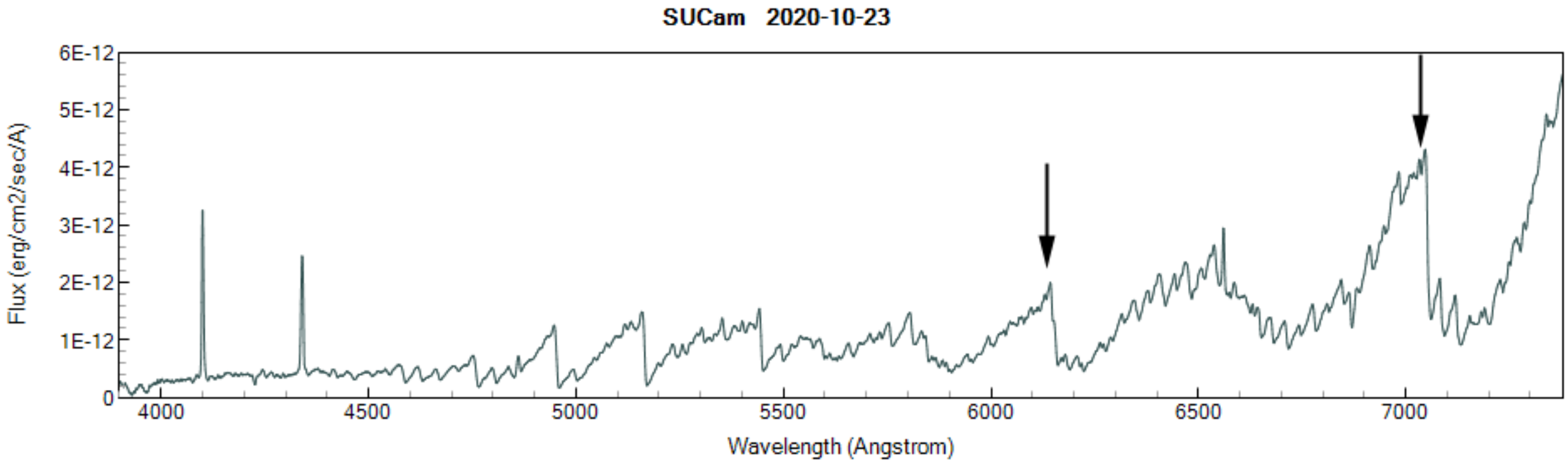

Figure 11. Position of two regions where the mean flux of each spectrum is measured.

In Figure 12 we plot the effective temperatures of all our SU Cam and RY Cep spectra against the ratio of these mean continuum fluxes. The narrowness of this distribution suggests that it may be possible to estimate effective temperature for Mira stars with spectral types between K4 and M8 by measuring this flux ratio. Fitting a 5$^{th}$ order polynomial to this distribution gives an R-squared of 0.98 and a rms residual of 45K.

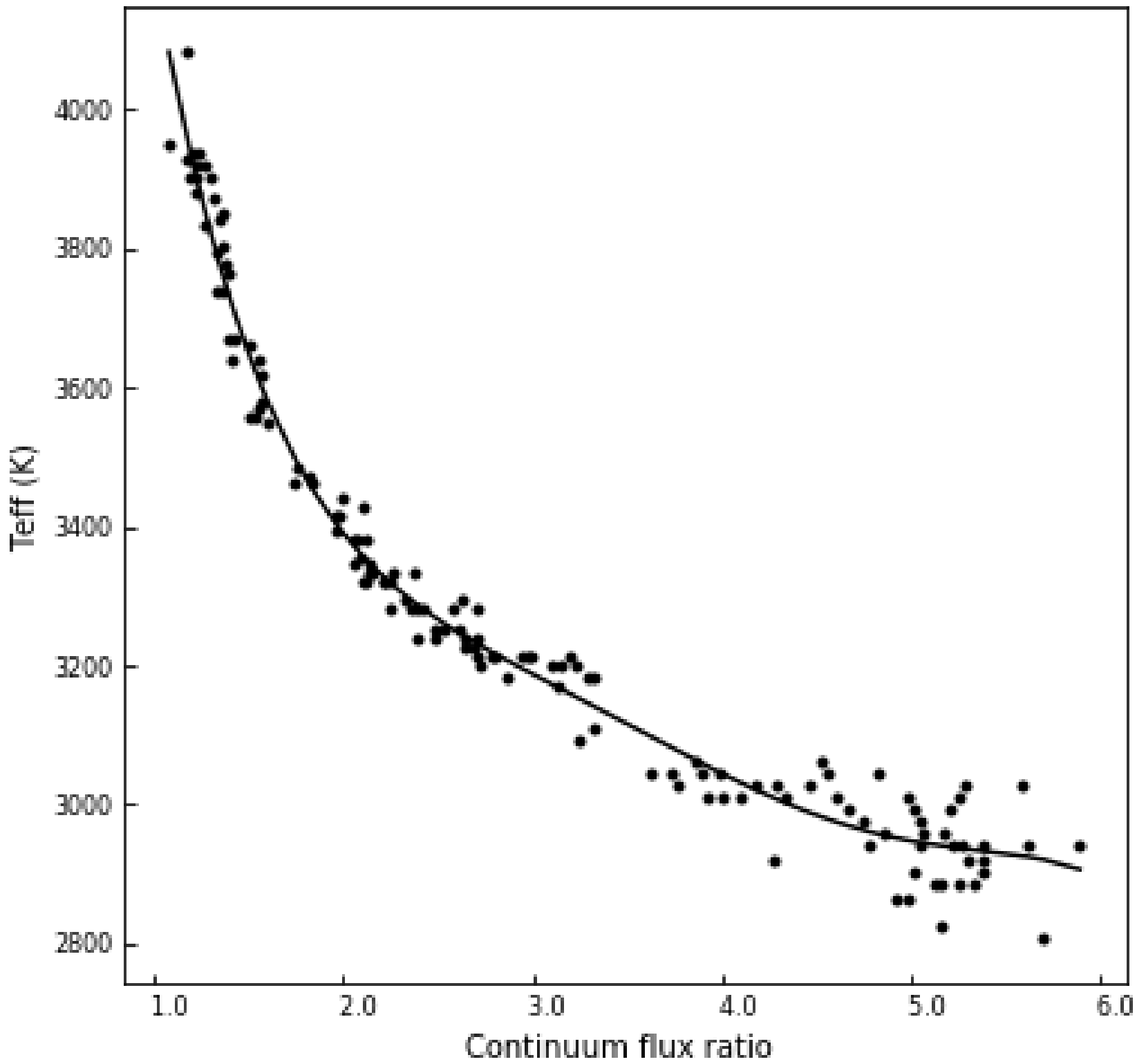

Figure 12. Effective temperature (Teff) vs continuum flux ratio for all SU Cam and RY Cep spectra plus a fitted 5$^{th}$ order polynomial.

## 9. Conclusions

This study shows that, with small telescopes suitably equipped and operated, it is possible using a combination of spectroscopy and photometry to monitor the behaviour of Mira stars such as SU Cam and RY Cep and to analyse how their brightness, spectra and Balmer emission varies over multiple pulsation cycles.

We found the following:

- during the time interval of this study the average pulsation periods of SU Cam and RY Cep were 292.49 days and 152.64 days respectively;
- there is a consistently different pattern of behaviour between the earlier spectral type RY Cep and the later SU Cam with flux in the B and V bands, flux in the Balmer emission lines and the effective temperature all peaking more sharply around the time of maximum brightness in SU Cam compared to RY Cep;
- maximum effective temperature coincides with maximum brightness in the V band in all pulsation cycles of both stars;
- there is a close correlation between effective temperature and V magnitude over the full brightness range in both rising falling branches of all cycles of both stars;
- the B-V colour index shows large variations over each cycle and from cycle to cycle and is a poor indication of effective temperature in these stars;
- relative Balmer line strengths, as measured by the Hα/Hδ flux ratio, change from decrement to increment as the spectral type becomes later with the transition occurring around spectral sub-type M3.6;
- the ratio of mean fluxes measured at two points near TiO molecular band heads can be used to estimate the effective temperature of a late K or M type Mira star.

## Acknowledgements

I acknowledge with thanks a constructive and helpful referee's report. I am grateful to Lee Anne Willson for her valuable comments and advice, and to John Percy for his helpful feedback. Richard Gray and Chris Corbally provided advice on MK standard stars and the latter gave useful comments on an early draft. This research made use of the AAVSO Photometric All-Sky Survey (APASS) and the AAVSO Variable Star Index (VSX). The software developed for this project made extensive use of the Astropy package and the efforts of many contributors to this valuable community resource are gratefully acknowledged.

Table 1. Stars from the Perkins catalogue used as standard spectral types for K0III to K5III.

| Spectral type | HD number | Common name |
|---|---|---|
| K0III | 197989 | epsilon Cyg |
| K1III | 037984 | 51 Ori |
| K2III | 054719 | tau Gem |
| K3III | 171443 | alpha Sct |
| K4III | 069267 | beta Cnc |
| K5III | 164058 | gamma Dra |

Table 2. Julian Date, cycle number, spectral sub-type, effective temperature (Teff) and Balmer emission line fluxes for each SU Cam spectrum.

| Julian Date of spectrum | Cycle | Spectral sub-type | Teff (K) | Hα line flux ($ergs/cm^2/sec$) | Hβ line flux ($ergs/cm^2/sec$) | Hγ line flux ($ergs/cm^2/sec$) | Hδ line flux ($ergs/cm^2/sec$) |
|---|---|---|---|---|---|---|---|
| 2457994.44120 | | M7.6 | 3045 | 9.99E-13 | 0.0 | 5.55E-13 | 1.86E-12 |
| 2458039.37220 | | M7.8 | 3011 | 3.90E-13 | 0.0 | 1.60E-13 | 4.58E-13 |
| 2458059.34370 | | M8.3 | 2922 | 2.29E-13 | 0.0 | 5.95E-14 | 2.27E-13 |
| 2458082.44680 | | M8.6 | 2865 | 1.35E-13 | 0.0 | 5.22E-14 | 0.0 |
| 2458115.43200 | | M8.9 | 2806 | 7.59E-14 | 0.0 | 0.0 | 0.0 |
| 2458137.48730 | 1 | M8.8 | 2826 | 5.79E-14 | 0.0 | 0.0 | 0.0 |
| 2458161.42780 | 1 | M8.5 | 2885 | 1.10E-13 | 0.0 | 0.0 | 0.0 |
| 2458174.42350 | 1 | M8.5 | 2885 | 1.17E-13 | 0.0 | 0.0 | 0.0 |
| 2458191.43750 | 1 | M7.6 | 3045 | 3.29E-13 | 0.0 | 2.92E-13 | 2.69E-13 |
| 2458212.42710 | 1 | M7.5 | 3061 | 7.40E-13 | 0.0 | 1.23E-12 | 9.87E-13 |
| 2458223.43220 | 1 | M6.6 | 3199 | 9.58E-13 | 9.12E-14 | 2.35E-12 | 2.89E-12 |
| 2458239.44960 | 1 | M6.4 | 3227 | 1.98E-12 | 2.70E-13 | 5.13E-12 | 6.57E-12 |
| 2458249.40210 | 1 | M6.0 | 3281 | 3.82E-12 | 7.22E-13 | 7.28E-12 | 8.50E-12 |
| 2458257.43280 | 1 | M5.7 | 3320 | 5.42E-12 | 1.22E-12 | 8.88E-12 | 9.74E-12 |
| 2458261.43060 | 1 | M5.6 | 3332 | 6.07E-12 | 1.49E-12 | 9.18E-12 | 1.00E-11 |
| 2458267.44220 | 1 | M5.4 | 3357 | 6.52E-12 | 1.91E-12 | 1.00E-11 | 1.04E-11 |
| 2458272.43560 | 1 | M5.7 | 3320 | 7.54E-12 | 1.78E-12 | 8.20E-12 | 8.49E-12 |
| 2458284.43990 | 1 | M6.0 | 3281 | 5.87E-12 | 9.64E-13 | 4.58E-12 | 3.87E-12 |
| 2458291.44420 | 1 | M6.3 | 3241 | 4.55E-12 | 6.09E-13 | 3.72E-12 | 3.30E-12 |
| 2458295.44880 | 1 | M6.6 | 3199 | 4.90E-12 | 4.22E-13 | 2.53E-12 | 2.24E-12 |
| 2458300.43730 | 1 | M6.5 | 3213 | 4.03E-12 | 3.12E-13 | 1.75E-12 | 1.38E-12 |
| 2458310.43950 | 1 | M6.7 | 3184 | 3.27E-12 | 2.38E-13 | 1.26E-12 | 9.78E-13 |
| 2458318.43060 | 1 | M7.3 | 3093 | 3.10E-12 | 2.12E-13 | 7.37E-13 | 4.74E-13 |
| 2458323.42060 | 1 | M7.8 | 3011 | 3.15E-12 | 1.75E-13 | 4.57E-13 | 4.95E-13 |
| 2458333.51160 | 1 | M7.7 | 3028 | 2.52E-12 | 1.21E-13 | 1.57E-13 | 2.54E-13 |
| 2458352.49640 | 1 | M7.8 | 3011 | 1.40E-12 | 0.0 | 0.0 | 0.0 |
| 2458379.45380 | 1 | M7.8 | 3011 | 4.31E-13 | 0.0 | 0.0 | 0.0 |
| 2458398.43130 | 1 | M7.9 | 2994 | 1.56E-13 | 0.0 | 0.0 | 0.0 |
| 2458414.39980 | 2 | M7.8 | 3011 | 7.56E-14 | 0.0 | 0.0 | 0.0 |
| 2458434.34880 | 2 | M7.7 | 3028 | 5.42E-14 | 0.0 | 0.0 | 0.0 |
| 2458477.41860 | 2 | M7.9 | 2994 | 1.05E-13 | 0.0 | 0.0 | 0.0 |
| 2458492.41150 | 2 | M8.3 | 2922 | 1.65E-13 | 0.0 | 0.0 | 0.0 |
| 2458512.40850 | 2 | M8.2 | 2941 | 1.83E-13 | 0.0 | 0.0 | 1.52E-13 |
| 2458519.35250 | 2 | M8.0 | 2977 | 2.08E-13 | 0.0 | 0.0 | 3.53E-13 |
| 2458533.33470 | 2 | M6.7 | 3184 | 5.86E-13 | 0.0 | 7.22E-13 | 2.00E-12 |
| 2458560.46490 | 2 | M6.0 | 3281 | 1.11E-12 | 0.0 | 2.66E-12 | 4.90E-12 |
| 2458575.36370 | 2 | M6.2 | 3254 | 9.04E-13 | 0.0 | 1.63E-12 | 3.80E-12 |
| 2458585.38330 | 2 | M6.3 | 3241 | 7.49E-13 | 0.0 | 1.02E-12 | 3.04E-12 |
| 2458595.44270 | 2 | M6.7 | 3184 | 7.75E-13 | 0.0 | 8.43E-13 | 2.56E-12 |
| 2458616.41010 | 2 | M7.7 | 3028 | 4.72E-13 | 0.0 | 3.68E-13 | 1.10E-12 |
| 2458643.43500 | 2 | M8.2 | 2941 | 3.04E-13 | 0.0 | 1.29E-13 | 1.15E-13 |
| 2458718.51510 | 3 | M8.5 | 2885 | 0.0 | 0.0 | 0.0 | 0.0 |
| 2458738.44610 | 3 | M8.4 | 2904 | 0.0 | 0.0 | 0.0 | 0.0 |
| 2458759.45040 | 3 | M7.6 | 3045 | 0.0 | 0.0 | 0.0 | 0.0 |
| 2458806.40450 | 3 | M8.1 | 2959 | 1.17E-13 | 0.0 | 0.0 | 8.28E-14 |
| 2458822.37100 | 3 | M7.7 | 3028 | 2.70E-13 | 0.0 | 2.06E-13 | 5.64E-13 |
| 2458840.42970 | 3 | M6.5 | 3213 | 6.74E-13 | 0.0 | 1.64E-12 | 3.68E-12 |
| 2458855.41260 | 3 | M6.5 | 3213 | 9.39E-13 | 0.0 | 2.42E-12 | 6.05E-12 |
| 2458864.38000 | 3 | M6.6 | 3199 | 9.79E-13 | 0.0 | 2.41E-12 | 6.55E-12 |
| 2458886.43030 | 3 | M6.6 | 3199 | 6.40E-13 | 0.0 | 2.04E-12 | 5.88E-12 |
| 2458900.46480 | 3 | M7.6 | 3045 | 6.21E-13 | 0.0 | 8.74E-13 | 3.39E-12 |
| 2458925.41210 | 3 | M8.0 | 2977 | 3.24E-13 | 0.0 | 2.34E-13 | 1.09E-12 |
| 2458934.37000 | 3 | M8.3 | 2922 | 2.76E-13 | 0.0 | 1.90E-13 | 7.87E-13 |
| 2458948.43670 | 3 | M8.1 | 2959 | 2.27E-13 | 0.0 | 8.00E-14 | 2.87E-13 |
| 2458955.37740 | 3 | M8.2 | 2941 | 1.80E-13 | 0.0 | 5.76E-14 | 1.66E-13 |
| 2458962.39400 | 3 | M8.1 | 2959 | 1.70E-13 | 0.0 | 5.00E-14 | 1.17E-13 |
| 2458976.42220 | 3 | M8.2 | 2941 | 1.33E-13 | 0.0 | 0.0 | 0.0 |
| 2458995.45310 | 3 | M7.8 | 3011 | 0.0 | 0.0 | 0.0 | 0.0 |
| 2459022.45080 | 4 | M7.7 | 3028 | 3.15E-14 | 0.0 | 0.0 | 0.0 |
| 2459048.43950 | 4 | M8.2 | 2941 | 1.67E-13 | 0.0 | 0.0 | 0.0 |
| 2459073.50590 | 4 | M7.8 | 3011 | 2.74E-13 | 0.0 | 0.0 | 0.0 |
| 2459098.46100 | 4 | M7.6 | 3045 | 6.03E-13 | 0.0 | 5.94E-13 | 1.25E-12 |
| 2459114.47510 | 4 | M6.5 | 3213 | 1.09E-12 | 0.0 | 2.80E-12 | 4.96E-12 |
| 2459131.45160 | 4 | M5.5 | 3345 | 3.47E-12 | 9.76E-13 | 1.05E-11 | 1.26E-11 |
| 2459146.37560 | 4 | M5.5 | 3345 | 6.51E-12 | 2.37E-12 | 1.52E-11 | 1.58E-11 |
| 2459164.42980 | 4 | M5.7 | 3320 | 8.27E-12 | 3.23E-12 | 1.39E-11 | 1.38E-11 |
| 2459179.32490 | 4 | M6.3 | 3241 | 8.10E-12 | 2.11E-12 | 7.54E-12 | 6.99E-12 |
| 2459196.39410 | 4 | M6.8 | 3170 | 7.25E-12 | 9.33E-13 | 2.60E-12 | 2.16E-12 |
| 2459214.35220 | 4 | M7.2 | 3109 | 4.59E-12 | 4.04E-13 | 7.70E-13 | 6.63E-13 |
| 2459230.33230 | 4 | M7.6 | 3045 | 3.09E-12 | 1.74E-13 | 2.84E-13 | 1.93E-13 |
| 2459249.47460 | 4 | M7.7 | 3028 | 1.60E-12 | 5.51E-14 | 0.0 | 0.0 |
| 2459258.40820 | 4 | M7.9 | 2994 | 1.21E-12 | 3.33E-14 | 0.0 | 0.0 |
| 2459264.41330 | 4 | M7.5 | 3061 | 8.19E-13 | 0.0 | 0.0 | 0.0 |
| 2459275.34950 | 4 | M8.5 | 2885 | 6.54E-13 | 0.0 | 0.0 | 0.0 |
| 2459291.42990 | 4 | M8.2 | 2941 | 2.63E-13 | 0.0 | 0.0 | 0.0 |
| 2459309.36300 | | M8.6 | 2865 | 1.10E-13 | 0.0 | 0.0 | 0.0 |
| 2459323.38720 | | M8.4 | 2904 | 1.51E-13 | 0.0 | 0.0 | 0.0 |
| 2459341.39820 | | M8.2 | 2941 | 1.51E-13 | 0.0 | 0.0 | 0.0 |

Table 3. Julian Date, cycle number, spectral sub-type, effective temperature (Teff) and Balmer emission line fluxes for each RY Cep spectrum.

| Julian Date of spectrum | Cycle | Spectral sub-type | Teff (K) | Hα line flux (ergs/cm$^2$/sec) | Hβ line flux (ergs/cm$^2$/sec) | Hγ line flux (ergs/cm$^2$/sec) | Hδ line flux (ergs/cm$^2$/sec) |
|---|---|---|---|---|---|---|---|
| 2458655.47320 | | M0.3 | 3882 | 7.12E-12 | 4.81E-12 | 3.98E-12 | 2.83E-12 |
| 2458665.45240 | | M0.8 | 3834 | 5.80E-12 | 3.07E-12 | 2.53E-12 | 1.74E-12 |
| 2458677.44190 | | M1.8 | 3738 | 4.53E-12 | 1.65E-12 | 1.29E-12 | 8.32E-13 |
| 2458690.45110 | | M3.6 | 3559 | 2.88E-12 | 7.07E-13 | 5.69E-13 | 3.46E-13 |
| 2458715.43760 | | M5.7 | 3320 | 6.41E-13 | 3.35E-14 | 0.0 | 0.0 |
| 2458732.37490 | 1 | M5.9 | 3294 | 2.26E-13 | 0.0 | 0.0 | 0.0 |
| 2458740.45250 | 1 | M6.2 | 3254 | 1.19E-13 | 0.0 | 0.0 | 0.0 |
| 2458753.36730 | 1 | M5.1 | 3393 | 2.24E-13 | 0.0 | 0.0 | 0.0 |
| 2458773.37670 | 1 | M2.5 | 3670 | 5.18E-12 | 2.75E-12 | 2.70E-12 | 1.47E-12 |
| 2458799.38690 | 1 | K5.6 | 3949 | 8.23E-12 | 6.21E-12 | 5.20E-12 | 3.62E-12 |
| 2458817.31610 | 1 | M0.1 | 3901 | 4.51E-12 | 1.97E-12 | 1.66E-12 | 1.13E-12 |
| 2458827.34330 | 1 | M1.2 | 3796 | 3.00E-12 | 9.18E-13 | 9.55E-13 | 5.47E-13 |
| 2458840.37610 | 1 | M1.8 | 3738 | 1.33E-12 | 3.52E-13 | 3.24E-13 | 1.38E-13 |
| 2458847.40540 | 1 | M3.6 | 3559 | 9.53E-13 | 1.62E-13 | 2.58E-13 | 8.71E-14 |
| 2458855.35170 | 1 | M4.5 | 3462 | 5.31E-13 | 0.0 | 0.0 | 0.0 |
| 2458864.31440 | 1 | M4.9 | 3417 | 2.88E-13 | 0.0 | 0.0 | 0.0 |
| 2458869.36570 | 1 | M5.6 | 3332 | 1.54E-13 | 0.0 | 0.0 | 0.0 |
| 2458886.38000 | 2 | M5.2 | 3381 | 1.15E-13 | 0.0 | 0.0 | 0.0 |
| 2458910.37200 | 2 | M2.8 | 3640 | 1.76E-12 | 7.63E-13 | 1.17E-12 | 6.49E-13 |
| 2458925.35070 | 2 | M0.1 | 3901 | 4.47E-12 | 2.98E-12 | 2.74E-12 | 1.66E-12 |
| 2458931.41790 | 2 | M1.1 | 3805 | 5.49E-12 | 3.80E-12 | 3.90E-12 | 2.77E-12 |
| 2458936.37410 | 2 | K5.7 | 3939 | 5.46E-12 | 4.83E-12 | 4.50E-12 | 3.00E-12 |
| 2458940.37870 | 2 | M0.6 | 3853 | 6.18E-12 | 4.71E-12 | 5.10E-12 | 3.53E-12 |
| 2458946.37370 | 2 | K5.9 | 3920 | 5.76E-12 | 4.58E-12 | 4.74E-12 | 3.13E-12 |
| 2458954.37210 | 2 | K5.8 | 3929 | 5.74E-12 | 3.88E-12 | 3.26E-12 | 1.87E-12 |
| 2458959.39670 | 2 | M0.4 | 3872 | 5.47E-12 | 2.43E-12 | 1.85E-12 | 1.15E-12 |
| 2458972.42060 | 2 | M1.4 | 3777 | 2.94E-12 | 7.44E-13 | 5.36E-13 | 1.82E-13 |
| 2458983.43620 | 2 | M3.4 | 3580 | 1.26E-12 | 2.25E-13 | 2.62E-13 | 0.0 |
| 2458994.42840 | 2 | M4.9 | 3417 | 8.81E-13 | 0.0 | 0.0 | 0.0 |
| 2459002.43200 | 2 | M4.8 | 3428 | 3.44E-13 | 0.0 | 0.0 | 0.0 |
| 2459015.43310 | 2 | M6.0 | 3281 | 2.07E-13 | 0.0 | 0.0 | 0.0 |
| 2459024.44080 | 2 | M6.5 | 3213 | 9.80E-14 | 0.0 | 0.0 | 0.0 |
| 2459041.44590 | 3 | M6.5 | 3213 | 6.47E-14 | 0.0 | 0.0 | 0.0 |
| 2459051.44400 | 3 | M5.9 | 3294 | 6.94E-14 | 0.0 | 0.0 | 0.0 |
| 2459058.44310 | 3 | M5.6 | 3332 | 1.13E-13 | 0.0 | 1.93E-13 | 0.0 |
| 2459093.37990 | 3 | M3.0 | 3620 | 1.44E-12 | 7.98E-13 | 1.87E-12 | 1.69E-12 |
| 2459102.44410 | 3 | M2.5 | 3670 | 3.10E-12 | 2.24E-12 | 4.20E-12 | 3.27E-12 |
| 2459112.38060 | 3 | M2.6 | 3660 | 3.36E-12 | 2.28E-12 | 3.46E-12 | 2.46E-12 |
| 2459120.39190 | 3 | M2.8 | 3640 | 2.32E-12 | 1.47E-12 | 2.26E-12 | 1.71E-12 |
| 2459129.43650 | 3 | M3.5 | 3569 | 1.49E-12 | 6.70E-13 | 1.24E-12 | 9.53E-13 |
| 2459140.39550 | 3 | M4.5 | 3462 | 7.20E-13 | 2.57E-13 | 6.25E-13 | 4.99E-13 |
| 2459149.34110 | 3 | M5.2 | 3381 | 4.15E-13 | 0.0 | 3.53E-13 | 2.67E-13 |
| 2459157.46540 | 3 | M5.7 | 3320 | 2.54E-13 | 0.0 | 1.62E-13 | 1.10E-13 |
| 2459172.31360 | 3 | M6.3 | 3241 | 1.30E-13 | 0.0 | 0.0 | 0.0 |
| 2459189.34830 | 4 | M5.6 | 3332 | 1.43E-13 | 0.0 | 0.0 | 0.0 |
| 2459203.38410 | 4 | M4.7 | 3440 | 1.96E-13 | 0.0 | 1.94E-13 | 9.44E-14 |
| 2459221.35100 | 4 | M3.7 | 3549 | 2.18E-12 | 1.20E-12 | 1.77E-12 | 1.11E-12 |
| 2459236.41120 | 4 | K5.7 | 3939 | 5.90E-12 | 5.91E-12 | 5.20E-12 | 3.54E-12 |
| 2459238.31480 | 4 | K5.9 | 3920 | 6.80E-12 | 6.17E-12 | 5.38E-12 | 3.46E-12 |
| 2459249.40320 | 4 | K4.2 | 4084 | 6.56E-12 | 6.40E-12 | 5.45E-12 | 3.62E-12 |
| 2459256.35220 | 4 | M0.1 | 3901 | 5.44E-12 | 4.30E-12 | 3.70E-12 | 2.79E-12 |
| 2459264.35450 | 4 | M0.7 | 3843 | 3.84E-12 | 2.18E-12 | 2.19E-12 | 1.43E-12 |
| 2459271.43120 | 4 | M1.5 | 3767 | 3.32E-12 | 1.25E-12 | 1.01E-12 | 6.73E-13 |
| 2459282.48690 | 4 | M4.4 | 3473 | 2.09E-12 | 3.42E-13 | 2.72E-13 | 1.63E-13 |
| 2459291.35470 | 4 | M4.3 | 3484 | 1.06E-12 | 1.93E-13 | 2.18E-13 | 5.54E-14 |
| 2459298.38450 | 4 | M5.2 | 3381 | 7.78E-13 | 0.0 | 0.0 | 0.0 |
| 2459308.46660 | 4 | M6.0 | 3281 | 5.79E-13 | 0.0 | 0.0 | 0.0 |
| 2459316.38150 | 4 | M6.2 | 3254 | 3.09E-13 | 0.0 | 0.0 | 0.0 |
| 2459322.38530 | 4 | M6.4 | 3227 | 2.70E-13 | 0.0 | 0.0 | 0.0 |
| 2459329.39380 | 4 | M6.5 | 3213 | 2.28E-13 | 0.0 | 0.0 | 0.0 |
| 2459339.40940 | | M6.0 | 3281 | 3.11E-13 | 0.0 | 1.15E-13 | 0.0 |

Table 4. Integrated flux emitted in each Balmer line and in the V band during each pulsation cycle in SU Cam.

| Cycle | Hα (ergs/cm$^2$) | Hβ (ergs/cm$^2$) | Hγ (ergs/cm$^2$) | Hδ (ergs/cm$^2$) | V band (ergs/cm$^2$) |
|---|---|---|---|---|---|
| 1 | 4.71E-05 | 6.49E-06 | 4.53E-05 | 4.80E-05 | 4.74E-03 |
| 2 | 8.17E-06 | 0.0 | 1.08E-05 | 2.55E-05 | 2.09E-03 |
| 3 | 7.41E-06 | 0.0 | 1.37E-05 | 3.90E-05 | 1.59E-03 |
| 4 | 6.72E-05 | 1.46E-05 | 7.68E-05 | 8.29E-05 | 5.06E-03 |

Table 5. Integrated flux emitted in each Balmer line and in the V band during each pulsation cycle in RY Cep.

| Cycle | Hα (ergs/cm$^2$) | Hβ (ergs/cm$^2$) | Hγ (ergs/cm$^2$) | Hδ (ergs/cm$^2$) | V band (ergs/cm$^2$) |
|---|---|---|---|---|---|
| 1 | 3.72E-05 | 2.09E-05 | 1.86E-05 | 1.19E-05 | 3.75E-03 |
| 2 | 3.10E-05 | 1.75E-05 | 1.72E-05 | 1.06E-05 | 3.74E-03 |
| 3 | 1.24E-05 | 7.01E-06 | 1.37E-05 | 1.06E-05 | 1.68E-03 |
| 4 | 2.93E-05 | 2.01E-05 | 1.89E-05 | 1.25E-05 | 3.08-03 |